\documentclass[twocolumn,amssymb,prl]{revtex4}
\usepackage{amsmath}
\usepackage[pdftex]{graphicx}
\usepackage{dcolumn}
\usepackage{bm}
\usepackage{caption}
\newcounter{one}
\pdfoutput=1

\begin{document}

\title{Programmable Multimode Quantum Networks}

\author
{*Seiji Armstrong$^{1,2,3}$, Jean-Fran\c cois Morizur$^{1,4}$, Jiri Janousek$^{1,2}$, Boris Hage$^{1,2}$, Nicolas Treps$^{4}$, Ping Koy Lam$^{2}$ and Hans-A. Bachor$^{1}$}

\affiliation{$^{1}$ARC Centre of Excellence for Quantum-Atom Optics, The Australian National University, Canberra, ACT 0200, Australia \\
$^{2}$Centre for Quantum Computation and Communication Technology,
Department of Quantum Science, The Australian National University,
Canberra, ACT 0200, Australia\\
$^{3}$Department of Applied Physics, School of Engineering, The University of Tokyo,
7-3-1 Hongo, Bunkyo-ku, Tokyo 113-8656, Japan\\
$^{4}$Laboratoire Kastler Brossel, Universit\'e Pierre et Marie Curie Paris 6, ENS, CNRS, Paris, France.}

\begin{abstract}

Entanglement between large numbers of quantum modes is the quintessential resource for future technologies such as the quantum internet. Conventionally the generation of multimode entanglement in optics requires complex layouts of beam-splitters and phase shifters in order to transform the input modes in to entangled modes. These networks need substantial modification for every new set of entangled modes to be generated. Here we report on the highly versatile and efficient generation of various multimode entangled states with the ability to switch between different linear optics networks in real time. By defining our modes to be combinations of different spatial regions of one beam, we may use just one pair of multi-pixel detectors each with M photodiodes in order to measure N entangled modes, with a maximum number of N=M modes. We program virtual networks that are fully equivalent to the physical linear optics networks they are emulating. We present results for N=2 up to N=8 entangled modes here, including N=2,3,4 cluster states. Our approach introduces flexibility and scalability to multimode entanglement, two important attributes that are highly sought after in state of the art devices.

\end{abstract}

\maketitle
*contact: seiji.armstrong@gmail.com

\clearpage

Multi-partite entanglement is not only of fundamental scientific interest, it is also the key ingredient for quantum information technologies \cite{Akira_book, Briegel, Menicucci_2006, Kimble_qnet}. In optics, several impressive demonstrations of multi-partite entanglement have been shown recently including an 8-photon cluster state \cite{8_photons} and a 9-mode state used for error correction \cite{Tokyo_error}. However, these schemes tend to employ one detection system per entangled mode/qubit, which introduces a lack of flexibility and is detrimental to its scalability. These optical setups are built to produce one set of outputs or to perform one given protocol; in order to change the output the optical hardware itself must be modified. We report here on a system with the ability to switch in real time between desired output states using just one detection scheme. 

Currently the well-established recipe for generating entanglement using continuous wave laser beams is to mix squeezed modes of light together at beam-splitters. It is possible to create $N$-mode entanglement given a network of N-1 beam-splitters with  N input modes, even with less than N squeezed modes \cite{one_squeezer}. In our scheme we co-propagate all possible spatial modes of light within one beam. Entanglement between co-propagating modes in one beam has been previously demonstrated with spatial modes \cite{Janousek, Ulrik}, and also in the frequency domain \cite{Pfister_cluster}. In the current work we radically expand the idea of one-beam entanglement by introducing the notion of emulating linear optics networks, by programming virtual networks that mix together different spatial regions of the light beam. These software based networks calculate the precise weighted combinations of the spatial regions required to emulate the physical networks. This is possible because the linear optical components in a typical network simply perform reversible operations, and can be represented by unitary matrices. It is worth stating explicitly that the entangled spatial modes that we produce are event-ready and unconditional {\em before} the detection process. The real-time virtual networks allow us to match the detection basis to the desired spatial mode basis contained within the beam, analogous to shaping a reference local oscillator beam.

\begin{figure*}[htb]
\centering
\includegraphics[width=16cm,clip]{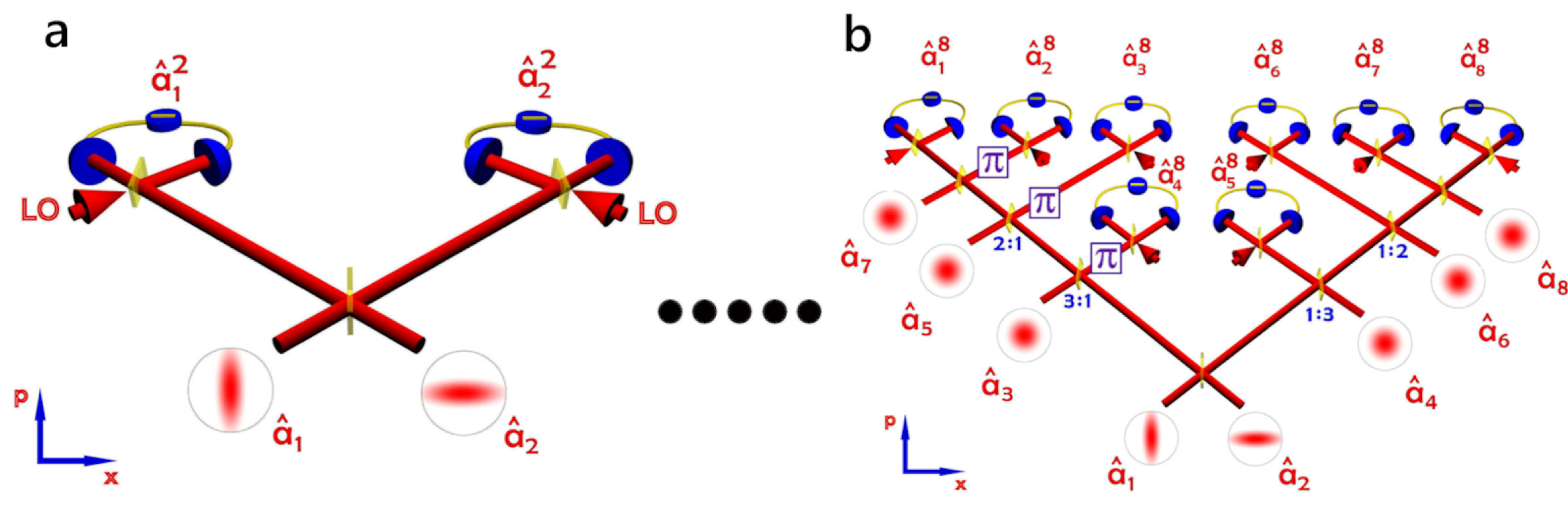} 
\caption{\textbf{Multimode entanglement via emulated linear optics networks.} Squeezed light and vacua are mixed together using unitary operations in order to produce entangled mode states. Unless otherwise stated beam-splitters are $50\%$ reflective. Superscripts denote mode basis and subscripts denote mode number. (\textbf{a}) The emulated linear optics network used to measure 2-mode EPR entanglement ($U_\mathrm{net}^\mathrm{2}$). (\textbf{b}) 8-mode entanglement via a calculated concatenation of beam-splitter and $\pi$ phase shift operations ($U_\mathrm{net}^\mathrm{8}$). The dots between \textbf{a} and \textbf{b} imply virtual networks for N = 3...7, not shown for brevity.
}
\label{FigVirtualNetworks}
\end{figure*}

Figure~\ref{FigVirtualNetworks} shows two such networks that produce a $2$-mode entangled state and an $8$-mode entangled state respectively. We also program virtual networks for $3$, $4$, $5$, $6$, and $7$-mode entangled states, the results of which are shown in Table 1 and Figure~\ref{FigNoise}. As a further demonstration of the versatility of our setup we produce linear $2$, $3$, and $4$ mode cluster states, which are highly entangled graph states garnering attention for their potential in quantum computing \cite{Briegel}. Table 2 summarises the cluster state measurements while Figure~\ref{FigInsep} shows that cluster states demand more stringent squeezing requirements than non-cluster inseparable states.\\

{\large \textbf{Results}}

\begin{figure*}[ht]
\centering
\includegraphics[width=10cm,clip]{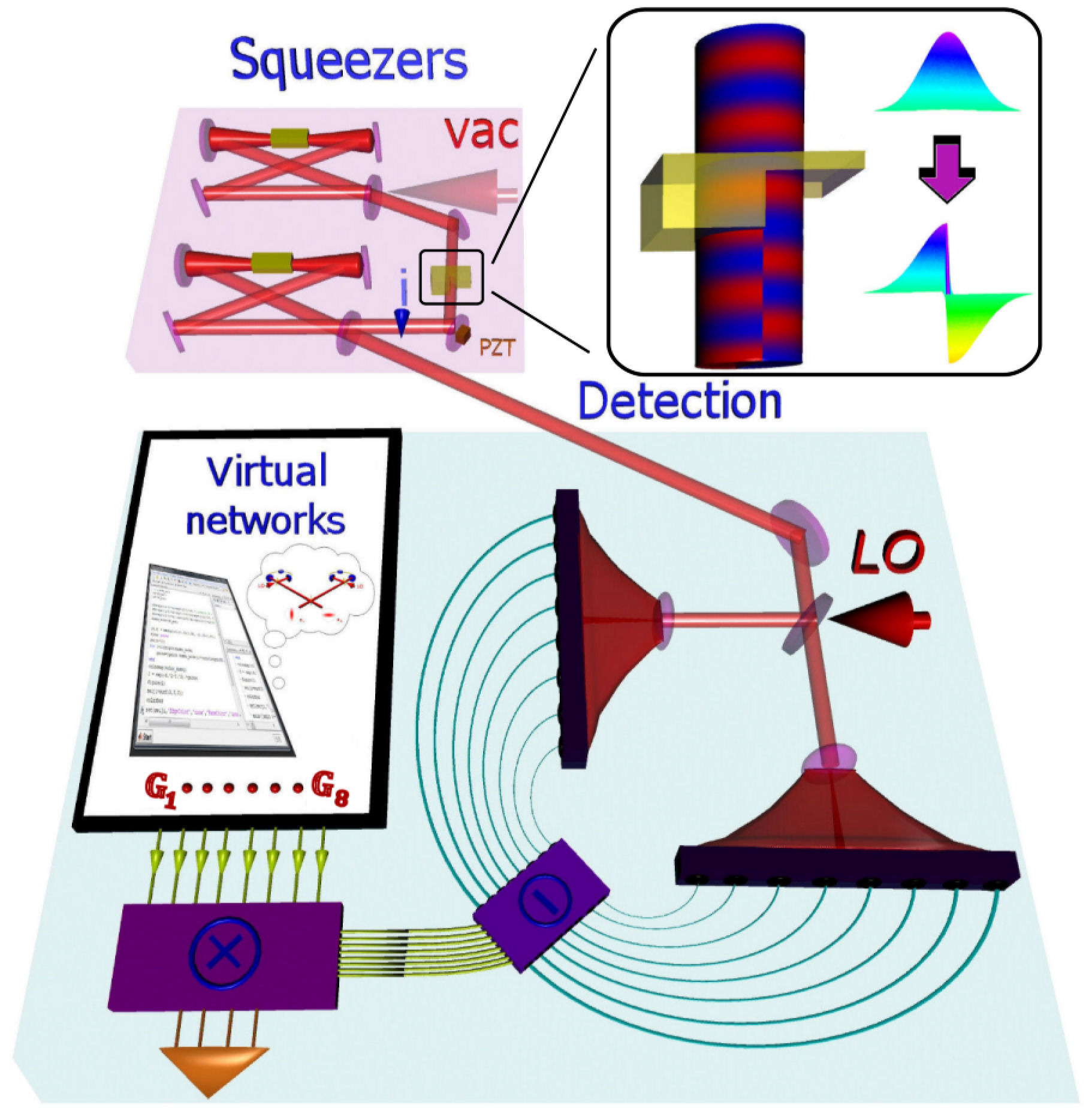} 
\caption{\textbf{Experimental setup.} (not to scale). Squeezed light is prepared and combined in \emph{squeezers} with a piezo electric transducer (PZT) controlling the phase between the two squeezed modes, locked in quadrature. Vacuum modes (vac) co-propagate so that the beam exiting \emph{squeezers} and entering \emph{detection} contains 8 measurable spatial modes. Multi-pixel homodyne detection (MPHD) is used to measure the quadrature amplitudes of the beam in 8 different regions, in \emph{detection}. Local oscillator (LO) gives a reference to phase quadratures. A PC is used to calculate electronic gain functions $G_{n}$ via the notion of \emph{virtual networks}. The detected beam is then projected onto a basis of measured modes (see equation 1). (\textbf{Inset}) Flip mode (FM) generation; half of the wave is phase retarded by half a wavelength, flipping the electric field amplitude.
}
\label{FigSetup}
\end{figure*}

\textbf{Measuring spatial modes.} By employing custom made multi-photodiode-homodyne-detectors (MPHD) that each contain an array of 8 photodiodes (see Figure~\ref{FigSetup}) we detect the light in 8 spatial regions and assign individual electronic gains to each spatial region. The linear combination of the 8 gain-adjusted photocurrents constitutes the measurement of one mode. 

More generally, we can express the measurement process of a complete set of spatial modes in one beam by the following:

\begin{eqnarray}
\mathbf{\hat{a}}&=&U\mathbf{\hat{i}}\\
&=&U_\mathrm{net} U_\mathrm{in}^\mathrm{N}\mathbf{\hat{i}}
\end{eqnarray}

where $\mathbf{\hat{a}}=(\hat{a}_{1},\cdots,\hat{a}_{N})^\mathrm{T}$ is the set of N measured modes projected by the N$\times 8$ unitary matrix $U$ acting on the 8 homodyne-subtracted photocurrent operators $\mathbf{\hat{i}}=(\hat{i}_{1},\cdots,\hat{i}_{8})^\mathrm{T}$. $U_\mathrm{in}^\mathrm{N}$ is an $N\times 8$ matrix made up of the top N rows of $U_\mathrm{in}$, the orthogonal $8\times 8$ unitary matrix that recovers the important set of 8 unmixed spatial modes that span the input basis (see Methods). Input modes are then mixed via $U_{net}$, which emulate linear optics networks, given by the $N\times N$ matrix:

\begin{equation}{U_\mathrm{net}} =
\left( {\begin{array}{cccc}
v_{1}^{1} &v_{2}^{1} &\cdots& v_{N}^{1}\\
v_{1}^{2} &v_{2}^{2} &\cdots& v_{N}^{2}\\
\vdots&\vdots&\ddots&\vdots\\
v_{1}^{N} & v_{2}^{N} &\cdots&v_{N}^{N}\\
 \end{array} } \right),
\label{eq:network}
\end{equation}

where $v_{p}^{n}\in\mathbb{R}$.

\begin{figure*}[ht]
\centering
\includegraphics[width=12cm,clip]{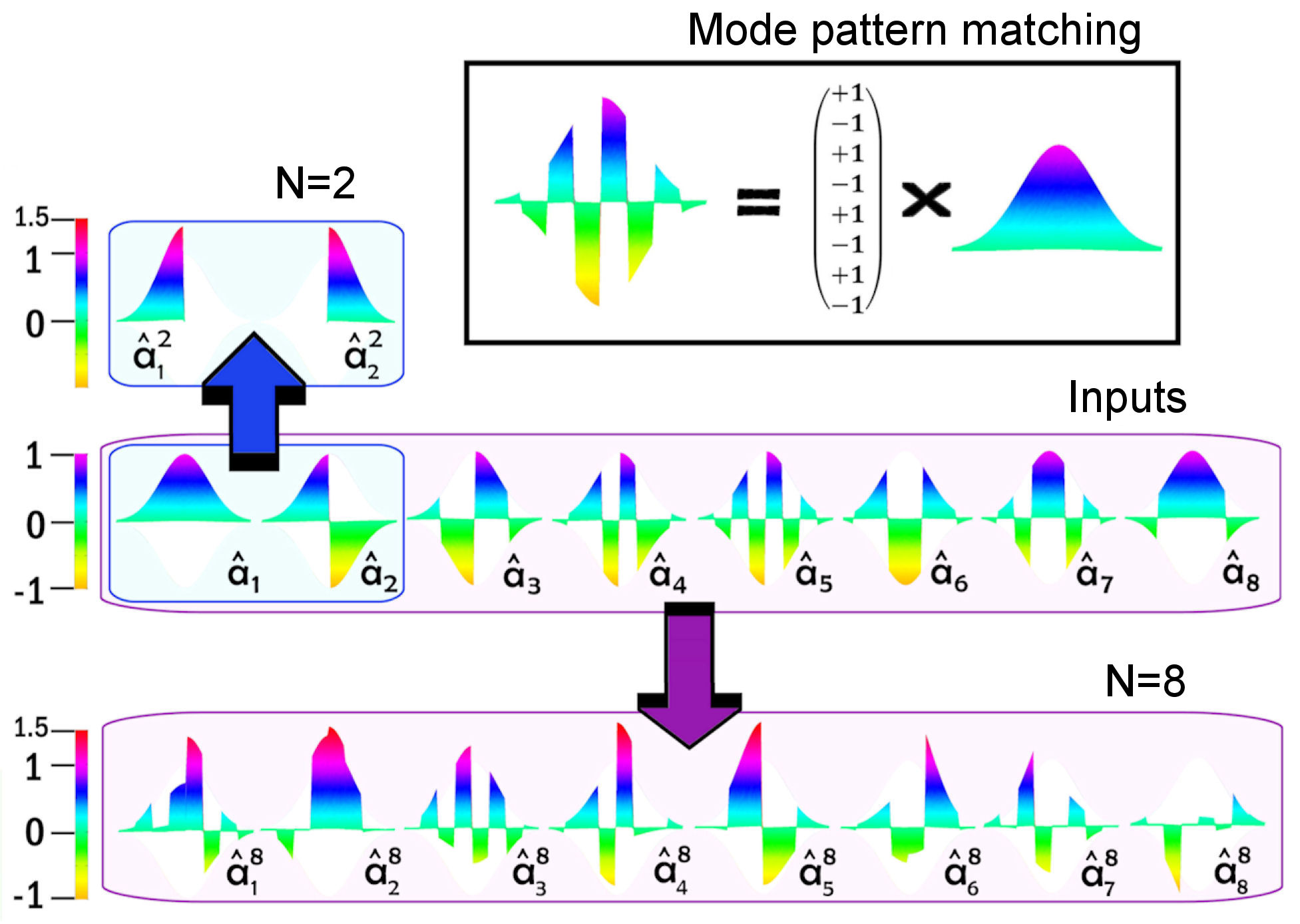} 
\caption{\textbf{Spatial mode patterns.} Measured modes are defined by spatial patterns of electric field amplitudes. Shown in the \emph{mode pattern matching} box is an example of how the spatial mode pattern for $\hat{a}_\mathrm{5}$ is matched by applying 8 electronic gain values ($G_\mathrm{5}$) to the detected Gaussian profile ($\mathbf{\hat{i}}$). The basis of input modes $\hat{a}_\mathrm{1}$...$\hat{a}_\mathrm{8}$ is shown in the middle row (see Methods). The arrows represent a mapping via the virtual networks $U_\mathrm{net}^\mathrm{2}$ and $U_\mathrm{net}^\mathrm{8}$ onto the respective bases of entangled modes; the top row shows the symmetric EPR or 2-mode basis, while the bottom row shows the 8-mode basis. There is a one-to-one correspondence between spatial mode bases shown here (\{$a_\mathrm{i}$\},\{$a_\mathrm{i}^\mathrm{2}$\}, \{$a_\mathrm{i}^\mathrm{8}$\}) and those shown in Figure~\ref{FigVirtualNetworks}. Again, spatial mode bases for $N$=3 to $N$=7 not shown for brevity.}
\label{FigSpatialModes}
\end{figure*}

\begin{figure*}[ht]
\centering
\includegraphics[width=10cm,clip]{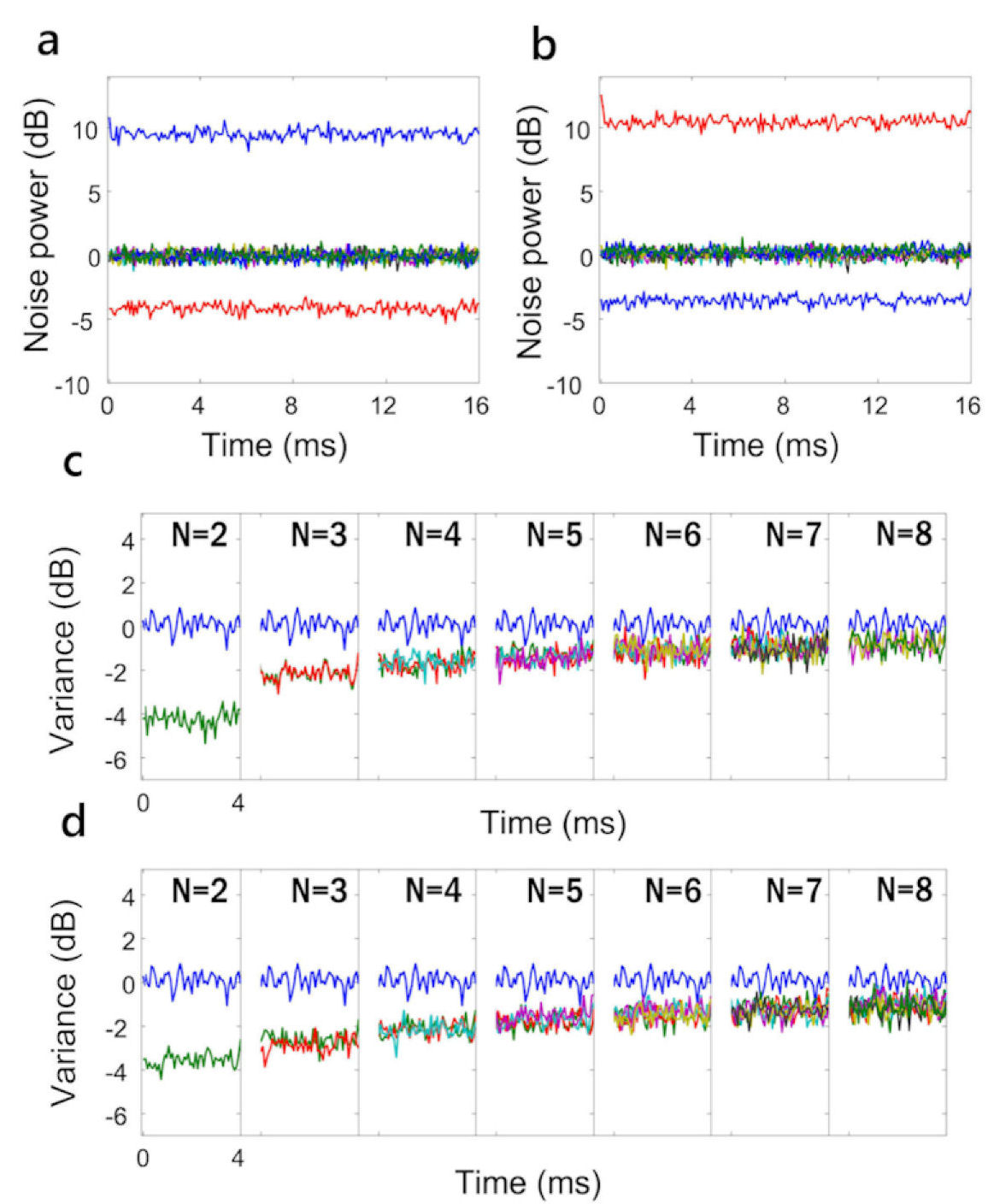} 
\caption{\textbf{Noise variance measurements of the spatial modes.} (\textbf{a}) x quadrature measurements of the input mode basis. The squeezed $\langle[\Delta x_\mathrm{1}]^{2}\rangle$ is shown in the red and anti-squeezed $\langle[\Delta x_\mathrm{2}]^{2}\rangle$ is shown in the blue. The x quadrature variances of the 6 vacua modes are measured to equal quantum noise (0dB). (\textbf{b}) p quadrature measurements. The anti-squeezed $\langle[\Delta p_\mathrm{1}]^{2}\rangle$ is shown in the red and squeezed $\langle[\Delta p_\mathrm{2}]^{2}\rangle$ is shown in the blue. The p quadrature variances of the 6 modes are again measured to equal quantum noise, confirming they are vacua. (\textbf{c}) These variances show the x quadrature correlations between modes as in the first half of the L.H.S. of equation (2) of the text. Every column shows N-1 traces of x quadrature correlations below shot noise, as well as the blue shot noise trace (0dB) normalised to two units of vacua. Each green trace shows $\langle [\Delta (\hat{x}_\mathrm{1}^\mathrm{N}-\hat{x}_\mathrm{2}^\mathrm{N})]^{2}\rangle$ for each N-mode basis. 
Each new colour represents the other N-1 variance correlation traces of equation (2). (\textbf{d}) Correlations between measured modes in p quadrature, second half of the L.H.S. of equation (2). Each green trace now shows $\langle [\Delta (\hat{p}_\mathrm{1}^\mathrm{N}+\hat{p}_\mathrm{2}^\mathrm{N}+g_\mathrm{3}\hat{p}_\mathrm{3}^\mathrm{N}+...+g_\mathrm{N}\hat{p}_\mathrm{N}^\mathrm{N})]^{2}\rangle$. The traces overlapping show that each pair of modes is entangled with the same strength as any other pair of modes, a result of optimising for symmetry in the virtual networks.
}
\label{FigNoise}
\end{figure*}

\begin{figure*}[ht]
\centering
\includegraphics[width=10cm,clip]{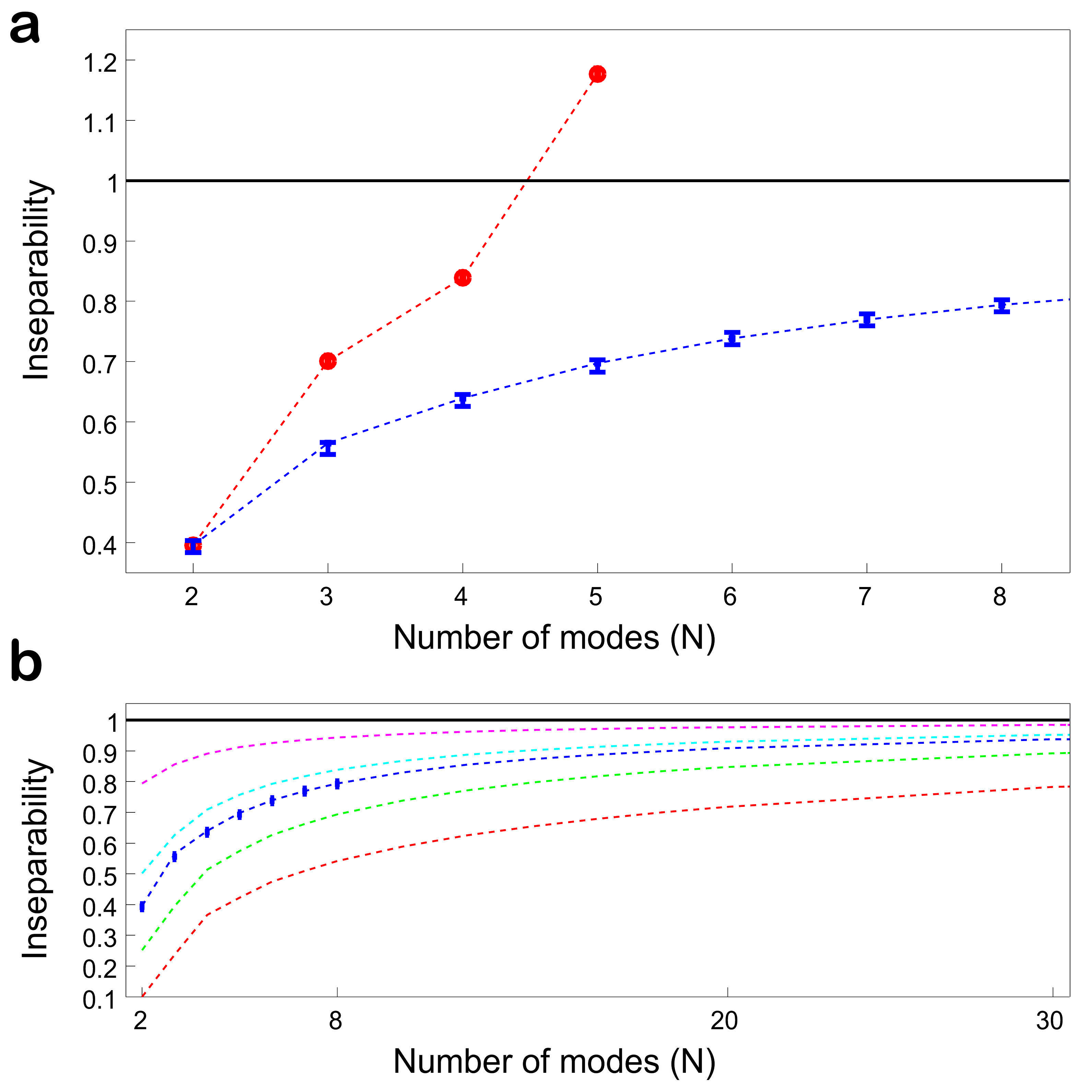} 
\caption{\textbf{Inseparability for different entangled mode bases.} The solid black line represents the bound of separability. Dashed lines represent theory. (\textbf{a}) The blue markers are the averaged measured experimental values for N-mode inseparability (right column of Table 1), and the dashed blue line joins the theoretical values of inseparability with the same two squeezed inputs used in the experiment. All experimental losses have been taken into account. The red circles are the measured experimental values for N-mode cluster states with theory indicated again by the dashed red line. Here the maximum value of each row in Table 2 is shown rather than the average value, in order to show that cluster states have a much more stringent requirement on squeezing levels (N=5 is clearly separable and not a cluster state). (\textbf{b}) All traces have two squeezed inputs and N-2 vacua modes, as in the experiment. What changes is the amount of squeezing in the two squeezed inputs, assumed here to be symmetric with equal anti-squeezing. From the top we have: -1dB (magenta); -3dB (cyan); experimental parameters (blue); experimental values (blue markers); -6dB (green); and -10dB (red).
}
\label{FigInsep}
\end{figure*}

This allows us to uniquely define a mode $\hat{a}_{n}$ by the 8 real numbers in the $n$th row of $U$, which we will label as the mode's gain vector $G_{n}$, such that $\hat{a}_{n}= G_\mathrm{n}\mathbf{\hat{i}}$. Therefore each spatial mode we measure, whether belonging to the input basis or an entangled mode basis, is defined by a unique pattern within the light beam. These spatial mode patterns, represented by Gaussian profiles modulated by respective electronic gains $G_{n}$, are shown visually in Figure~\ref{FigSpatialModes}, while the detection stage of Figure~\ref{FigSetup} shows how we implement this experimentally. The spatial modes are orthogonal to each other, spanning a basis so that the independent measurement of each mode is possible \cite{Beck,Beck2}.\\

\textbf{Input basis.} We create two amplitude squeezed modes via optical parametric amplification (OPA). The first mode is converted to a flip mode (FM) by phase delaying half its beam by half a wavelength, $\pi$ (see inset  of Figure~\ref{FigSetup}). The FM is overlapped in quadrature with the Gaussian mode (GM) output of the second OPA upon reflection of its output coupler \cite{Delaubert}. These two squeezed modes are the first two modes of what we refer to as the input basis; $\hat{a}_\mathrm{1}$ and $\hat{a}_\mathrm{2}$. Six co-propagating vacua modes are measured by calculating $G_{n}$ vectors that are orthogonal to both $\hat{a}_\mathrm{1}$ and $\hat{a}_\mathrm{2}$. These vacua modes (labelled $\hat{a}_\mathrm{3}$...$\hat{a}_\mathrm{8}$) complete the input mode basis (see middle row of Figure~\ref{FigSpatialModes}). Measuring these modes amounts to matching the detection basis by following equation (1) and setting $U_\mathrm{net}=I$ (see Methods). 

Each spatial mode is characterised by the continuous-variable (CV) quadrature operators $\hat{x}$ and $\hat{p}$ of the electric field operator. The $\hat{x}$ and $\hat{p}$ variance measurements of the eight modes in the input basis are shown in Figure~\ref{FigNoise}a,b. Here, $\langle[\Delta x_\mathrm{GM}]^{2}\rangle=\langle[\Delta x_\mathrm{1}]^{2}\rangle=-4.3\pm 0.05$dB and $\langle[\Delta p_\mathrm{FM}]^{2}\rangle=\langle[\Delta p_\mathrm{2}]^{2}\rangle=-3.7\pm 0.05$dB below the standard quantum noise, and the variances of the vacua are verified to equal quantum noise.\\

\textbf{Entangled mode bases.} Programming a virtual network amounts to calculating the precise expression for $U_{net}$. The unitaries we have access to in programming the virtual networks are beam-splitters and $\pi$ phase shifts. $U_{net}$ is the concatenation of all of these unitaries that make up a linear optics network. The $\pi$ phase shift is equivalent to multiplying $\hat{a}$ by $-1$. Note that arbitrary phase shifts are forbidden as each measurement naturally corresponds to detection at a fixed phase defined by a shared reference beam, the local oscillator. Importantly, optimal virtual networks are calculated allowing for optimisation of beam-splitters due to asymmetries in the squeezing levels of input modes.

The most intuitive virtual network we create is the 2-mode EPR state \cite{EPR} shown in Figure~\ref{FigVirtualNetworks}a with 2 squeezed inputs. Here we engineer spatial mode patterns which have no spatial overlap; the left half of the beam is entangled with the right half (see the top left of Figure~\ref{FigSpatialModes}). Entangled modes belonging to other bases share spatial overlap but are nevertheless spatially orthogonal. 

Spatial modes measured in an entangled mode basis are given a superscript N to distinguish them from modes in the input basis; $a_\mathrm{1}^\mathrm{2}$ and $a_\mathrm{2}^\mathrm{2}$ represent the two modes spanning the N=2-mode EPR basis. See the Methods section for details on how we create virtual networks for each of the N=2, 3, 4, 5, 6, 7, and 8-mode bases. In general we construct networks pertaining to N modes by concatenating N-1 virtual beam-splitters with vacua on unused input ports. This is a highly efficient approach to creating multimode entanglement as the arduous tasks of mode matching and alignment are replaced with the ease of programming.\\

\textbf{Cluster states.} Attracting attention for their potential in one-way quantum computing schemes, cluster states are a type of highly entangled Gaussian graph state \cite{Cluster_PVL,Cluster_Comp}. They satisfy the quadrature relation $(\hat{p}_{a}-\sum_{b\in N_\mathrm{a}} \hat{x}_{b})\rightarrow 0$. As infinite squeezing would require infinite energy and are thus unrealisable, one is limited to the production of approximate cluster states in the laboratory, and there have been demonstrations of up to four-mode CV cluster states thus far \cite{Tokyo_cluster,China_cluster}.

In order to measure cluster states in one beam we must be able to access the correct quadratures of each entangled mode. Here we measure $2$, $3$, and $4$ mode linear cluster states, however measuring arbitrary cluster shapes would require modifying the optical setup (see Methods).\\

{\large \textbf{Discussion}}

In order to verify entanglement between measured modes we use the well-established van Loock-Furusawa inseparability criteria \cite{PVL_AF}. For an $N$-mode entangled state, it is sufficient to satisfy $N-1$ inseparability inequalities:
\begin{equation}
\begin{array}{l c c l l r}
\textbf{(I)}\\
\langle [\Delta (\hat{x}_\mathrm{1}-\hat{x}_\mathrm{2})]^{2}\rangle\\
+\langle [\Delta (\hat{p}_\mathrm{1}+\hat{p}_\mathrm{2}+g_\mathrm{3}\hat{p}_\mathrm{3}+...+g_\mathrm{N}\hat{p}_\mathrm{N})]^{2}\rangle < 1,\\
\cdots\\
\cdots\\
\textbf{(N-1)}\\ 
\langle [\Delta (\hat{x}_\mathrm{N-1}-\hat{x}_\mathrm{N})]^{2}\rangle\\
+\langle [\Delta (g_\mathrm{1}\hat{p}_\mathrm{1}+...+g_\mathrm{N-2}\hat{p}_\mathrm{N-2}+\hat{p}_\mathrm{N-1}+\hat{p}_\mathrm{N})]^{2}\rangle < 1.
\end{array}
\end{equation}
with free parameters $g_{i}$, to be optimised for maximum inseparability. We have omitted the superscript N here for clarity, as the above holds for any mode basis. The subscripts $n$ of $\hat{x}$ and $\hat{p}$ here indicate the $n$th mode in the N-mode basis. Table 1 summarises the measured degrees of inseparability for all $N-1$ inequalities in each $N$-mode basis, given in roman numerals. Table 2 summarises the more stringent inseparability required for unweighted cluster states (all homodyne gains \{$g_{i}$\} are set to $1$). The relevant inseparability inequalities satisfy the cluster state quadrature relationship written in the form of equation (4), and are written out explicitly in the Methods for reference.

The terms in equation (4) measure the degree of correlations between any two modes in a given basis. For the modes to be inseparable each of these correlation variances (correlations in the $x$ quadrature and anti-correlations in the $p$ quadrature) must be in the quantum regime, that is below the normalised quantum noise of two units of vacua. Figure~\ref{FigNoise}c,d shows this to be the case in our experimental measurements. Although we are limited here to 8 modes due to our detection scheme, this scheme is scalable to higher numbers of mode entanglement even without increasing the number of squeezing resources, as shown in the simulation traces of Figure~\ref{FigInsep}. As we increase the simulated number of modes in the basis up to 30, the degree of inseparability approaches the classical bound of 1 due to the vacuum noise penalty for each additional unsqueezed mode input. Entanglement is shown to hold here however, even with current squeezing levels. Importantly, there is no loss incurred during the transformation of the squeezed input modes into a set of entangled modes, as can be seen by the agreement of the theoretical predictions and the experimental values of Figure~\ref{FigInsep}a. This equates to perfect mode matching at every virtual beam-splitter. Figure~\ref{FigInsep}b explores how inseparability scales with different squeezing levels. Measuring a larger number of inseparable modes experimentally requires only an increase in the number of photodiodes in the MPHD, and importantly no modification of the optical setup. Note that this is not true for cluster states, and the number of squeezed inputs must be increased accordingly.

For the special case of $N=2$, optimal EPR entanglement \cite{Reid} is measured to be $0.58 \pm 0.01$. Optimising for the beam-splitter reflectivity \cite{biased,biased_kate} we find that due to the slight asymmetry between input squeezing levels, the optimal beam-splitter ratio here is not $50\%$, but rather $48.8\%$, leading to a very slight improvement over the symmetric network. Each unique beam-splitter reflectivity changes the mapping of $U_\mathrm{net}$ such that formally the beam of light contains an infinite number of mode bases. The versatility of our scheme comes from being able to match the detection basis to a network that has been optimised for an arbitrary set of inputs.

The entanglement demonstrated in the current work allows for such protocols as quantum teleportation \cite{one_squeezer, teleportation, Akira_teleportation}. To perform complex protocols such as one-way measurement based quantum computations \cite{Briegel} modifications are needed. We have shown with the current setup it is possible to create cluster states, the resource state for one-way quantum computations. To perform computations on the cluster states however we need access to arbitrary homodyne angles of each mode, and the ability to perform feed-forward to any desired mode \cite{UPMC_classical}. Both are feasible with existing technologies \cite{UPMC_quantum,Jeff_thesis} as discussed in the supplementary material.\\

Emulating linear optics networks by mixing copropagating spatial modes is a highly efficient method for generating multimode entanglement. Otherwise arduous and potentially lossy tasks such as mode matching during the construction of a linear optics network are performed effortlessly and losslessly via software controlled combinations of the spatial modes. We have shown that although correlations weaken if more squeezing resources are not added, (non-cluster) entangled modes scale here as the number of orthogonal modes measurable within the beam. The maximum number of measurable modes corresponds directly to the number of photodiodes in each pair of the multi-pixel detectors. We have demonstrated this by measuring N=2, 3, 4, 5, 6, 7, and 8-mode entanglement within one beam, including up to 4 mode cluster states, switching between them in real time. The ability to perform a wide range of protocols and optimise networks for asymmetry using just one optical setup offers versatility to future networks that will utilise entanglement as a resource. \\

\begin{table}[h]
\caption{Inseparability of entangled modes based on the van Loock-Furusawa criteria. Each row shows that for a basis of N quantum modes, the N-1 values obtained from quadrature variances are well below 1. This verifies entanglement of the N modes.}
\begin{tabular*}{0.5\textwidth}{@{\extracolsep{\fill}}lcccccccr}\textbf{N} & \textbf{I} & \textbf{II} & \textbf{III} & \textbf{IV} & \textbf{V} & \textbf{VI} & \textbf{VII} & \textbf{Avg.} \\\hline\textbf{2} & 0.39 & & & & & & & 0.39 \\\textbf{3} & 0.56 & 0.56 & & & & & & 0.56 \\\textbf{4} & 0.64 & 0.63 & 0.64 & & & & & 0.64\\\textbf{5} & 0.69 & 0.69 & 0.70 & 0.70& & & & 0.69\\\textbf{6} & 0.73 & 0.73 & 0.75 & 0.74 & 0.74 & & & 0.74\\\textbf{7} & 0.77 & 0.78 & 0.77 & 0.76 & 0.77 & 0.77 & & 0.77\\\textbf{8} & 0.79 & 0.79 & 0.78 & 0.81 & 0.79 & 0.80 & 0.79 & 0.79\\ \hline  \end{tabular*}\\ 
\label{table}
\:*Uncertainty is $\pm 0.01$ in all cases.
\end{table}

\begin{table}[h]
\caption{Inseparability of cluster states.}
\begin{tabular*}{0.5\textwidth}{@{\extracolsep{\fill}}lccccr}\textbf{N} & \textbf{I} & \textbf{II} & \textbf{III} & \textbf{IV} & \textbf{Avg.} \\\hline\textbf{2} & 0.39 & & & & 0.39 \\\textbf{3} & 0.49 & 0.70 & & & 0.59 \\\textbf{4} & 0.79& 0.67 & 0.84 & &  0.76\\\textbf{5} & 0.79 & 0.67 & 1.10 & 1.18&  0.93\\ \hline  \end{tabular*}\\ 
\label{table2}
\:*Uncertainty is $\pm 0.01$ in all cases.
\end{table}

\textbf{Methods}

{\footnotesize \textbf{Experimental setup}. We use a dual-wavelength continuous-wave Nd:YAG laser at 1064 nm and 532 nm. The optical parametric amplifiers (OPA) each contain a periodically poled KTP crystal in a bow-tie cavity. The squeezed beams are almost identical in purity, with squeezing levels of approximately -6 dB and anti-squeezing of 8.5 dB. The beam containing the 8 spatially orthogonal modes (see main text) is made highly elliptical in order to be measured by the MPHD, which has a linear array of 8 photodiodes. The photodiode array used is a Hamamatsu InGaAs PIN photodiode array (G7150) which actually has 16 photodiodes however we choose to use only 8 of these in the present experiment. The filling factor for the array is $90\%$, meaning that $10\%$ of the light does not hit an active surface. The quantum efficiency for the photodiodes are $80\%$.\\

\textbf{Virtual Networks.} For even numbered mode bases ($N=2,4,6,8$) the method for creating the virtual network is as follows. The two squeezed modes $a_1$ and $a_2$ are combined on a half reflecting beamsplitter (HBS). As the output of this HBS is an EPR state we choose to call this the EBS.
The EBS outputs are symmetrically combined with $N-2$ vacua, as in Figure 1 of the main text. The BSs are then given by $\hat{B}(cos^{-1}1/\sqrt{\frac{N}{2}-n})$, where $n$ is the number of BSs between the EBS and the BS in question. For N=4 and N=6 and N=8, mode output 2 is swapped with mode output N-1. For N=8, an additional swap of output modes 4 and 5 is made. For odd numbered mode bases ($N=3,5,7$) the method is the same with the following modifications. The EBS has its reflectivity changed to $r=\frac{1}{2}-\frac{1}{2N}$.  (See for example references \cite{GHZ_Tokyo, tritter} for more details on $N=3$). The vacua are mixed using beamsplitters as above, with one output arm having one less vacuum input. $\pi$ phase shifts are applied to all BS outputs on the left of the EBS except for the one left output exiting the last BS. Mode outputs 1 and N-1 are swapped, and the network for N=7 has an additional swap between output modes 3 and 4. The homodyne gains $g_\mathrm{i}$ are optimised using a genetic algorithm in order to maximally satisfy the van Loock-Furusawa inequalities. These gains $g_\mathrm{i}$ scale the contributions of the quadrature variances and are independent from calculations regarding $U_\mathrm{net}$. Here, optimal homodyne gains are calculated using two measures: minimising the mean of the $N-1$ inequalities; and minimising the variance of the set of inequalities. A trade-off between the two measures is needed, and preference is given to minimising the mean of the inequalities.\\

\textbf{Spatial mode bases.} The input matrix is defined as follows:

\begin{equation}{U_\mathrm{in}} =\frac{1}{\sqrt{8}}
\left( {\scriptsize\begin{array}{cccccccc}
1& 1& 1& 1& 1& 1& 1& 1\\
1 &1& 1 &1 &-1 &-1 &-1 &-1\\
1 &1 &-1 &-1 &1 &1 &-1 &-1\\
-1 &1 &1 &-1& 1 &-1 &-1 &1\\
1 &-1& 1 &-1 &1 &-1 &1 &-1\\
-1& 1 &1 &-1 &-1 &1 &1 &-1\\
-1 &1 &-1 &1 &1 &-1 &1 &-1\\
-1& -1 &1 &1 &1 &1 &-1 &-1\\
 \end{array} } \right)
\label{eq:network_in}
\end{equation}

with $U_\mathrm{in}^{*}U_\mathrm{in}=I$. Each row of $U_\mathrm{in}$ represents the 8 electronic gains that match the detection basis to the input modes. For example the top row containing all ones recovers the standard Gaussian $TEM_{00}$ mode (GM), and the second row recovers the phase-flipped Gaussian mode (FM). By setting $U_\mathrm{net}=I$ we can label each row of $U_\mathrm{in}$ as $G_\mathrm{n}^\mathrm{in}$. Formally, $U_\mathrm{in}^{N}=(I_\mathrm{N}$  $O_\mathrm{N,(8-N)}) U_\mathrm{in}$, where $O_\mathrm{N, (8-N)}$ is a zero matrix of size N by (8-N). 

The linear optics network for the ideal and symmetric 2-mode EPR basis is simply a HBS:

\begin{equation}{U_\mathrm{net}^\mathrm{2}} ={B(\frac{1}{\sqrt{2}})} =
\left( {\footnotesize\begin{array}{cc}
\frac{1}{\sqrt{2}}&\frac{1}{\sqrt{2}}\\
\frac{1}{\sqrt{2}}&-\frac{1}{\sqrt{2}}
 \end{array} } \right)
\label{eq:network_2}
\end{equation}

From equation (2) we get:
\begin{eqnarray*}{\hat{a}_\mathrm{1}^\mathrm{2}\choose \hat{a}_\mathrm{2}^\mathrm{2}} &=&
\frac{1}{\sqrt{8}}U_\mathrm{net}^\mathrm{2}
U_\mathrm{in}^\mathrm{2}
\mathbf{\hat{i}}\\
&=&{\textstyle \frac{1}{\sqrt{8}}}\left( {\scriptsize\begin{array}{cc}
\frac{1}{\sqrt{2}}&\frac{1}{\sqrt{2}}\\
\frac{1}{\sqrt{2}}&-\frac{1}{\sqrt{2}}\\
 \end{array} } \right)
\left( {\scriptsize\begin{array}{cccccccc}
1& 1& 1& 1& 1& 1& 1& 1\\
1 &1& 1 &1 &-1 &-1 &-1 &-1\\
 \end{array} } \right)
\mathbf{\hat{i}}\\
&=&{\textstyle \frac{1}{\sqrt{8}}}\left( {\scriptsize\begin{array}{cccccccc}
\sqrt{2}& \sqrt{2}&\sqrt{2}&\sqrt{2}&0&0&0&0\\
0&0&0&0&\sqrt{2}& \sqrt{2}&\sqrt{2}&\sqrt{2}\\
 \end{array} } \right)
\mathbf{\hat{i}}\\
&=&{\hat{G}_\mathrm{1}^\mathrm{2}\choose \hat{G}_\mathrm{2}^\mathrm{2}} \mathbf{\hat{i}},\\
\label{eq:EPR_network}
\end{eqnarray*}

from which we see that indeed $\hat{G}_\mathrm{1}^\mathrm{2}$ and $ \hat{G}_\mathrm{2}^\mathrm{2}$ indeed share no part of the detected light. We show this ideal EPR basis in Figure~\ref{FigSpatialModes} in order to emphasise the spatial separation. Note that the factor $\frac{1}{\sqrt{8}}$  has been omitted from the scale in Figure~\ref{FigSpatialModes} for clarity.\\

The optimised network uses a beam-splitter reflectivity of $48.8\%$, and produces the following output modes:

\begin{eqnarray*}{\hat{a}_\mathrm{1}^\mathrm{2,opt}\choose \hat{a}_\mathrm{2}^\mathrm{2,opt}} &=&
\frac{1}{\sqrt{8}}U_\mathrm{net}^\mathrm{2,opt}
U_\mathrm{in}^\mathrm{2}
\mathbf{\hat{i}}\\
&=&{\textstyle \frac{1}{\sqrt{8}}}\left( {\scriptsize\begin{array}{cc}
0.699&0.716\\
0.716&-0.699\\
 \end{array} } \right)
\left( {\scriptsize\begin{array}{cccccccc}
1& 1& 1& 1& 1& 1& 1& 1\\
1 &1& 1 &1 &-1 &-1 &-1 &-1\\
 \end{array} } \right)
\mathbf{\hat{i}}\\
&=&{\textstyle \frac{1}{\sqrt{8}}}\left( {\tiny\begin{array}{cccccccc}
1.414&1.414&1.414&1.414&-0.017&-0.017&-0.017&-0.017\\
0.017&0.017&0.017&0.017&1.414&1.414&1.414&1.414\\
 \end{array} } \right)
\mathbf{\hat{i}}\\
&=&{\hat{G}_\mathrm{1}^\mathrm{2,opt}\choose \hat{G}_\mathrm{2}^\mathrm{2,opt}} \mathbf{\hat{i}}.\\
\label{eq:EPR_network}
\end{eqnarray*}

\textbf{Measuring cluster states.} The difference between a two-mode cluster state and a two-mode EPR state is a Fourier transform on one mode. The Fourier transform is a rotation of $\frac{\pi}{2}$ degrees:  $F=R(\pi/2)=\bigl(\begin{smallmatrix}
cos\pi/2&-sin\pi/2\\sin\pi/2&co\pi/2
\end{smallmatrix} \bigr)=\bigl(\begin{smallmatrix}
0&-1\\1&0
\end{smallmatrix} \bigr)$. Therefore we get $F\bigl(\begin{smallmatrix}
\hat{x}\\
\hat{p}
\end{smallmatrix} \bigr)=\bigl(\begin{smallmatrix}
-\hat{p}\\
\hat{x}
\end{smallmatrix} \bigr)$.  It follows that the homodyne measurements we perform in the 2-mode cluster basis \{$p_\mathrm{1}-x_\mathrm{2},p_\mathrm{2}-x_\mathrm{1}\}_{cluster}$ and the 2-mode EPR basis \{$x_\mathrm{1}-x_\mathrm{2},p_\mathrm{1}+p_\mathrm{2}\}_{EPR}$ are equivalent. Therefore we may perform local Fourier transforms so long as we can match the homodyne detection basis for individual modes. It is important to note that this convenient basis change will not always be possible for different clusters. However, by shaping the local oscillator we may have access to arbitrary cluster states within the one-beam. This was out of the scope for the current experiment.

The criteria for verifying the measurements of the various cluster states are given below \cite{Cluster_PVL,Tokyo_cluster} with the results summarised in Table 2: \\

\begin{equation}
\begin{array}{l c c l l r}
\textbf{N=2}\\
\textbf{I} \langle [\Delta (\hat{p}_\mathrm{1}-\hat{x}_\mathrm{2})]^{2}\rangle+\langle [\Delta (\hat{p}_\mathrm{2}-\hat{x}_\mathrm{1})]^{2}\rangle < 1,\\
\end{array}
\end{equation}
\begin{equation}
\begin{array}{l c c l l r}
\textbf{N=3}\\
\textbf{I} \langle [\Delta (\hat{p}_\mathrm{1}-\hat{x}_\mathrm{2})]^{2}\rangle+\langle [\Delta (\hat{p}_\mathrm{2}-\hat{x}_\mathrm{1}-\hat{x}_\mathrm{3})]^{2}\rangle < 1,\\
\textbf{II} \langle [\Delta (\hat{p}_\mathrm{2}-\hat{x}_\mathrm{1}-\hat{x}_\mathrm{3})]^{2}\rangle+\langle [\Delta (\hat{p}_\mathrm{3}-\hat{x}_\mathrm{2})]^{2}\rangle < 1,\\
\end{array}
\end{equation}
\begin{equation}
\begin{array}{l c c l l r}
\textbf{N=4}\\
\textbf{I} \langle [\Delta (\hat{p}_\mathrm{1}-\hat{x}_\mathrm{2})]^{2}\rangle+\langle [\Delta (\hat{p}_\mathrm{2}-\hat{x}_\mathrm{1}-\hat{x}_\mathrm{3})]^{2}\rangle < 1,\\
\textbf{II} \langle [\Delta (\hat{p}_\mathrm{2}-\hat{x}_\mathrm{1}-\hat{x}_\mathrm{3})]^{2}\rangle+\langle [\Delta (\hat{p}_\mathrm{3}-\hat{x}_\mathrm{2}-\hat{x}_\mathrm{4})]^{2}\rangle < 1,\\
\textbf{III} \langle [\Delta (\hat{p}_\mathrm{3}-\hat{x}_\mathrm{2}-\hat{x}_\mathrm{4})]^{2}\rangle+\langle [\Delta (\hat{p}_\mathrm{4}-\hat{x}_\mathrm{3})]^{2}\rangle < 1,\\
\end{array}
\end{equation}
\begin{equation}
\begin{array}{l c c l l r}
\textbf{N=5}\\
\textbf{I} \langle [\Delta (\hat{p}_\mathrm{1}-\hat{x}_\mathrm{2})]^{2}\rangle+\langle [\Delta (\hat{p}_\mathrm{2}-\hat{x}_\mathrm{1}-\hat{x}_\mathrm{3})]^{2}\rangle < 1,\\
\textbf{II} \langle [\Delta (\hat{p}_\mathrm{2}-\hat{x}_\mathrm{1}-\hat{x}_\mathrm{3})]^{2}\rangle+\langle [\Delta (\hat{p}_\mathrm{3}-\hat{x}_\mathrm{2}-\hat{x}_\mathrm{4})]^{2}\rangle < 1,\\
\textbf{III} \langle [\Delta (\hat{p}_\mathrm{3}-\hat{x}_\mathrm{2}-\hat{x}_\mathrm{4})]^{2}\rangle+\langle [\Delta (\hat{p}_\mathrm{4}-\hat{x}_\mathrm{3}-\hat{x}_\mathrm{5})]^{2}\rangle < 1,\\
\textbf{IV} \langle [\Delta (\hat{p}_\mathrm{4}-\hat{x}_\mathrm{3}-\hat{x}_\mathrm{5})]^{2}\rangle+\langle [\Delta (\hat{p}_\mathrm{5}-\hat{x}_\mathrm{4})]^{2}\rangle < 1.\\
\end{array}
\end{equation}

}

\textbf{Acknowledgements}

{\footnotesize
This research was conducted by the Australian Research Council Centre of Excellence for Quantum-Atom Optics (project number CE0348178), in collaboration with the network "High-dimensional entangled systems" (HIDEAS FP7-ICT-221906) funded by the European Union, as well as the Australian Research Council Centre of Excellence for Quantum Computation and Communication Technology (project number CE110001029). The authors thank Akira Furusawa and the Furusawa group for discussions. SA is grateful for funding from the Australia-Asia Prime Minister`s Award. BH acknowledges funding from the Alexander von Humboldt foundation.}\\

\textbf{Author contributions}

{\footnotesize
S.A., J-F.M., J.J., N.T. and H-A.B. designed the experiment.  J.J. designed and built both of the optical parametric amplifiers. S.A., J-F.M. and J.J. constructed and performed the experiment. B.H. designed and built the multi-photodiode homodyne detectors, taught S.A. how to code the digital locking system, and designed the software filters used in data analysis. P.K.L. and H-A.B. supervised the experiment. S.A. calculated and optimised the virtual networks, performed the data analysis and wrote the manuscript. J-F.M. wrote the data acquisition code and provided support in the data analysis.}\\

\textbf{Competing financial interests}

{\footnotesize
The authors declare no competing financial interests. A signed form is attached.}\\

\end{document}